\documentclass{article}

\usepackage[preprint, nonatbib]{neurips_2026}

\usepackage[utf8]{inputenc}
\usepackage[T1]{fontenc}
\usepackage{amsmath,amssymb}
\usepackage{graphicx}
\usepackage{booktabs}
\usepackage{xcolor}
\usepackage[numbers,sort&compress]{natbib}
\usepackage[colorlinks=true,linkcolor=blue,citecolor=blue,urlcolor=blue]{hyperref}

\title{How Many Tools Should an LLM Agent See?\\ A Chance-Corrected Answer}

\author{%
  Vyzantinos Repantis \quad Ameya Gawde \quad Harshvardhan Singh \quad Joey Blackwell II \\
  Meta Platforms
}

\begin{document}
\maketitle

\begin{abstract}
Before an LLM agent can use a tool, a retrieval system must decide which candidate
tools to show to the agent. How long should that shortlist be? Show too many tools
and the model struggles to choose. Show too few and the correct tool may not appear.
Most systems apply a fixed shortlist size to every query, but no standard metric
exists to evaluate whether that size was appropriate. We treat the number of tools
shown to an LLM agent as the object of evaluation and we apply Bits-over-Random
(BoR)~\citep{repantis2026bor}, a chance-corrected metric that asks whether success
at a given depth is better than what random selection would achieve at that same
depth. We evaluate BoR across three tool-selection benchmarks, multiple scorers, and
registries ranging from 20 to 3{,}251 tools. We then turn the same principle into a
reinforcement learning (RL) reward for choosing tool shortlist depth per query. The
RL agent is deliberately simple, serving as a probe of the metric rather than a
proposed system. As the shortlist grows, random chance of including the correct tool
rises, so the reward naturally decreases, reducing the need for an engineered depth
penalty. On BFCL (370 tools), the learned policy nearly matches the coverage of
showing 50 tools ($90.3\%$ vs $90.8\%$) while presenting only 7 on average. On
ToolBench (3{,}251 tools), a fixed shortlist of 5 tools achieves higher aggregate
coverage ($64.7\%$ vs $61.9\%$) but finds nothing on hard queries (correct tool
ranked 6th--20th). The BoR agent finds $16.7\%$ on those same queries by searching
deeper. Downstream validation with Claude Sonnet 4.6 indicates that shorter adaptive
lists also improve the LLM's ability to select the right tool: $93.1\%$ versus
$87.1\%$ when always shown 5 tools, widening to $76.8\%$ vs $60.9\%$ on
medium-difficulty queries where the correct tool is present but not ranked first.
\end{abstract}


\section{Introduction}
Every LLM agent framework faces the same design choice: how many tools should the
model see per query? This search depth $K$ (the number of tools shown to the model)
is typically chosen once and never revisited.

Most practitioners choose a fixed $K$ ($K{=}5$, $K{=}10$, or ``show everything'') and
apply it uniformly to every query regardless of difficulty. This is common in
deployed RAG pipelines, MCP servers, and LLM agent frameworks. However, fixed $K$
leaves the depth-selection question unanswered: it cannot distinguish easy queries,
where one tool is enough, from harder queries. No standard methodology exists to tell
the builder whether the chosen $K$ was appropriate. Document retrieval has
depth-aware metrics such as nDCG@K and MAP@K, but tool selection has no equivalent.

Tool registries are small (typically 50 to 500 items), but showing too many tools is
costly. At roughly 200 tokens per tool description, a shortlist of 100 candidates
consumes 20K tokens before the query is even processed. Evaluating depth requires
separating three questions: Was the right tool presented? Did the LLM choose it? Did
execution succeed? Without this separation, it is difficult to diagnose where a
system fails. This paper focuses on the first question.

To evaluate search depth, we apply the Bits-over-Random (BoR)
metric~\citep{repantis2026bor}. BoR measures a system's success rate relative to
random chance: $\mathrm{BoR} = \log_2(P_{\mathrm{obs}}/P_{\mathrm{rand}})$, where
$P_{\mathrm{obs}}$ is the system's observed success rate and $P_{\mathrm{rand}}$ is
the rate expected from random selection at the same depth. This structure also yields
a reward signal. Because $P_{\mathrm{rand}}$ grows as more tools are shown, the reward
naturally decreases with depth. The RL agent, thus, learns how many tools to present
without requiring a task-specific depth penalty beyond a small constant continuation
cost.

We evaluate BoR across nine conditions spanning three tool-selection benchmarks
(BFCL, MetaTool, ToolBench), three scorer types, and registries from 20 to 3{,}251
tools. In one case, BoR flags a broken scorer that other metrics miss. We also
validate BoR on three document retrieval benchmarks (SciFact, NFCorpus, MS MARCO) in
Appendix~\ref{app:retrieval}.

Against Fixed-K baselines, an RL agent trained with BoR achieves comparable coverage
at substantially lower depth in most conditions. On BFCL (370 real tools), it
achieves $90.3 \pm 2.4\%$ coverage at $K{=}7.4$, nearly matching FK=50 ($90.8\%$) at
$7\times$ less depth. On ToolBench (3{,}251 tools), Fixed $K{=}5$ achieves higher
aggregate coverage ($64.7\%$ vs $61.9 \pm 0.6\%$), but finds nothing on hard queries
where the BoR agent finds $16.7 \pm 4.3\%$ by searching deeper ($K{=}5.7$).

\paragraph{Scope.} This paper evaluates and controls search depth, focusing on
whether the correct tool is shown to the LLM. A downstream validation
(Section~\ref{sec:downstream}) indicates that adaptive depth also improves tool
choice, but execution correctness is out of scope. The RL agent is a probe of the
metric's properties, not a proposed production architecture.

\section{Related Work}

\subsection{Adaptive Retrieval Depth and Context Sizing}
Prior work on how many items to present to an LLM differs in the signal used to
choose depth.

\paragraph{Heuristic and score-distribution methods.} Adaptive candidate sizing
predates the LLM era: Kratzwald and Feuerriegel~\citep{kratzwald2018adaptive} showed
that a fixed number of retrieved documents can hurt deep QA performance. More recent
work picks $K$ per query using patterns in the retrieval scores: Taguchi et
al.~\citep{taguchi2025adaptivek} select depth via the largest gap in sorted
similarity scores, while Xu et al.~\citep{xu2025cluster} use score clustering. These
methods are efficient but assume the scorer is reliable. When the scorer itself is
weak, score patterns offer little basis for choosing depth.

\paragraph{Supervised and learned selection.} Iratni et
al.~\citep{iratni2025dynamic} train a predictor of how many contexts to retrieve and
integrate it into a RAG pipeline. Their results show that retrieving fewer contexts
can still hurt generation when relevant evidence is omitted. DPS~\citep{meng2025dps}
uses supervised fine-tuning to select a variable-size subset of passages rather than
a fixed top-K.

\paragraph{Binary retrieve-or-not decisions.} Rather than choosing how many items to
retrieve, some systems focus on whether to retrieve at all.
SmartRAG~\citep{gao2025smartrag} uses reinforcement learning to decide whether
retrieval is needed for a given query. Self-RAG~\citep{asai2024selfrag} lets the model
decide mid-generation whether to pause and retrieve, and
Adaptive-RAG~\citep{jeong2024adaptiverag} routes queries to different retrieval
strategies based on estimated complexity. FLARE~\citep{jiang2023flare} retrieves when
the model's own confidence drops during generation. These systems adapt whether to
retrieve but treat depth $K$ as fixed once triggered.

\paragraph{RL for adaptive depth.} While the systems above decide whether to
retrieve, reinforcement learning offers a way to learn how much to retrieve.
DynamicRAG~\citep{sun2025dynamicrag} is closely related: it trains a reranker to
dynamically adjust both the order and the number of retrieved documents per query,
using downstream generation quality as the reward. It focuses on documents rather
than tools and optimizes downstream task performance rather than selectivity. Related
truncation work addresses the same question from a ranking perspective: Arampatzis et
al.~\citep{arampatzis2009} use score distribution modeling to find optimal cutoff
points in ranked lists. Choppy~\citep{bahri2020choppy} applies a transformer over
relevance scores. Similarly, Meng et al.~\citep{meng2024ranked} study truncation
specifically for LLM-based re-ranking pipelines. These methods study truncation and
cutoff selection, but not with a chance-corrected depth signal. Tool registries pose
a different challenge: every additional candidate costs tokens and risks confusing
the model, yet the pool is small enough that random success is non-negligible at
modest $K$.

\subsection{Tool Selection, Routing, and Presentation}
Tool-selection evaluation is less mature than document retrieval evaluation. Shi et
al.~\citep{shi2025toolsavvy} show that standard retrieval models perform notably worse
on tool retrieval than on document retrieval, suggesting the problem has distinct
properties. Standard benchmarks (ToolBench~\citep{qin2024toolllm},
MetaTool~\citep{huang2024metatool}, BFCL~\citep{patil2025bfcl}, among
others~\citep{guo2024stabletoolbench,li2023apibank,yao2024taubench}) mainly evaluate
downstream tool-use performance rather than whether the search depth itself was
appropriate. For instance, BFCL often evaluates function calling with only a few
candidate functions per query. MetaTool evaluates tool-usage awareness across
21{,}127 queries but leaves search depth fixed. ToolSandbox~\citep{lu2025toolsandbox}
studies robustness in the presence of distraction tools, rather than evaluating depth
efficiency. Meanwhile, LongFuncEval~\citep{kate2025longfunceval} and Rabinovich and
Anaby-Tavor~\citep{rabinovich2025robustness} show that function-calling accuracy
degrades as tool catalogs grow or as semantically similar tools are added. But
neither evaluates whether the search depth itself was appropriate for a given query.

Several systems address dynamic tool filtering and routing.
Less-is-More~\citep{paramanayakam2025lessismore} dynamically reduces the number of
tools shown to LLMs using similarity-based filtering, reporting up to $70\%$ lower
execution time on edge hardware. MemTool~\citep{lumer2025memtool} manages tool context
across multi-turn conversations through agent autonomy or deterministic rules.
ToolRerank~\citep{zheng2024toolrerank} applies rule-based adaptive truncation.
RAG-MCP~\citep{gan2025ragmcp} demonstrates clear degradation as $N$ grows from 1 to
11{,}100 MCP tools, highlighting the need for depth control.
DTDR~\citep{patel2025dtdr} conditions tool retrieval on both the query and evolving
execution history, reporting improvements in retrieval metrics and downstream
function-calling success. These systems improve which tools are retrieved or how they
are filtered, but they do not treat search depth itself as the primary object of
learning or evaluation.

\subsection{RL Reward Design for Retrieval and Tool Use}
Reinforcement learning (RL) reward design for retrieval and tool use is an active
area. The main challenge is credit assignment: when the final answer is wrong, it is
often unclear whether retrieval failed or the model failed to use what it retrieved.

\paragraph{Reward design for tool RL.} ToolRL~\citep{qian2025toolrl} systematically
explores reward types, scales, and granularity for tool learning, finding that
rewarding only the final outcome does not give the agent enough signal to improve its
tool choices. ARTIST~\citep{singh2025artist} combines agentic reasoning with
reinforcement learning using outcome-based rewards.

\paragraph{Process-level and multi-factor rewards for retrieval.} Rather than
rewarding only the final answer, several systems reward intermediate retrieval steps
directly. ProRAG~\citep{wang2026prorag} adds step-level supervision to prevent the
agent from reaching correct answers through flawed reasoning.
EVO-RAG~\citep{ji2025curriculum} uses a multi-factor step-level reward covering
relevance, redundancy, and efficiency, reducing retrieval depth while improving
multi-hop QA. InForage~\citep{qian2025inforage} and Zhang et
al.~\citep{zhang2025process} also explore intermediate rewards for retrieval quality.

\paragraph{Information-theoretic principles in reward design.}
InfoRM~\citep{miao2024inform} applies information-theoretic principles to reward
modeling for RLHF alignment, introducing an information bottleneck to filter spurious
features. The approach applies information theory to reward design, but for alignment
rather than retrieval or depth.

These papers explore what reward structure works, but they rely on accuracy-based,
outcome-based, or multi-factor heuristic rewards. None use a chance-corrected metric
as the reward for depth control.

\subsection{Information-Theoretic and Chance-Corrected Metrics}
SePer~\citep{dai2025seper} measures retrieval utility by the reduction in an LLM's
semantic perplexity. MIGRASCOPE~\citep{zheng2026migrascope} evaluates retriever
quality with mutual-information and divergence-based analyses. Song et
al.~\citep{song2026lessismore} use information gain as a pruning criterion for RAG
context selection. These metrics evaluate retrieval quality through information
content, but they are not chance-corrected against a random baseline and have not been
used as RL reward signals for depth control.

Chance-corrected selectivity metrics have a longer history in cheminformatics.
Truchon and Bayly~\citep{truchon2007bedroc} introduced BEDROC, a Boltzmann-enhanced
discrimination metric for virtual screening. The enrichment factor has been standard
in drug discovery for decades~\citep{huang2006benchmarking}. Despite the generality
noted by Truchon and Bayly~\citep{truchon2007bedroc}, these metrics have not been
adopted by the IR or NLP communities. BoR brings the same chance-corrected
perspective to LLM retrieval, connecting two literatures that developed independently.

We are not aware of prior work that treats search depth as a first-class property of
tool retrieval and learns it with a chance-corrected reward.

\section{Method}

\subsection{Bits-over-Random}
BoR was introduced as a chance-corrected selectivity metric in Repantis et
al.~\citep{repantis2026bor}. For a selection task with corpus size $N$, where $R_q$
items are relevant to a query and $K$ items are presented, the random baseline is the
probability of including at least one relevant item by chance. The hypergeometric
distribution gives:
\begin{equation}
P_{\mathrm{rand}} = 1 - \frac{\binom{N-R_q}{K}}{\binom{N}{K}}.
\label{eq:prand}
\end{equation}
For single-tool queries ($R_q{=}1$), this simplifies to $P_{\mathrm{rand}} = K/N$. The
observed success rate $P_{\mathrm{obs}}$ is the fraction of queries where at least one
relevant item appears in the top $K$ (Success@K). BoR applies to other success rules
as well (e.g., Recall@K), but all experiments in this paper use the $\geq 1$ rule.
BoR, measured in bits, is the log-ratio:
\begin{equation}
\mathrm{BoR} = \log_2\!\left(\frac{P_{\mathrm{obs}}}{P_{\mathrm{rand}}}\right).
\label{eq:bor}
\end{equation}
Each bit represents one doubling of selectivity over random chance. $\mathrm{BoR}{=}0$
means the system performs at random level. $\mathrm{BoR}{=}10$ means $1{,}024\times$
better than random. $\mathrm{BoR}<0$ means worse than random.

When $P_{\mathrm{obs}}{=}1$ (perfect success), the maximum achievable selectivity is
$\mathrm{BoR}_{\max}(K) = -\log_2(P_{\mathrm{rand}}(K))$. When $R_q$ is unknown, the
optimistic ceiling $\mathrm{BoR}_{\mathrm{opt}}(K) = \log_2(N/K)$ assumes $R_q{=}1$
for all queries and provides an upper bound on achievable selectivity for any system
on the same corpus~\citep{repantis2026bor}. This ceiling makes BoR self-calibrating
across systems: it depends only on corpus size and depth, not on the retrieval method.

A key property of BoR is the doubling rule~\citep{repantis2026bor}: when success rates
plateau ($P_{\mathrm{obs}}$ stops improving with depth), doubling $K$ costs
approximately 1 bit of selectivity, since $\Delta\mathrm{BoR} \approx
-\log_2(K_2/K_1)$. Because $P_{\mathrm{obs}}$ is bounded by 1, maintaining selectivity
while increasing depth becomes impossible once $P_{\mathrm{obs}}>0.5$. This is the
mathematical basis for the self-pruning property (Section~\ref{sec:selfpruning}).

For tool selection the same structure applies, but $N$ is small (typically 50--500),
so $P_{\mathrm{rand}}$ rises quickly with $K$.

\subsection{Markov Decision Process (MDP) Formulation}
We frame depth selection as an MDP where one query constitutes one episode. Given a
user query, a scorer (BM25 or a sentence-embedding model) ranks all $N$ candidate
tools by estimated relevance. The RL agent examines this ranked list one tool at a
time. After each step, it observes the current scores and decides: present the tools
examined so far to the LLM (STOP), or look at the next candidate (CONTINUE).

\paragraph{State.} At each step $t$, the RL agent observes the similarity scores of
the tools examined up to step $t$ (including the top score, the gap between the first
and current score, and the score spread), the current depth $k_t$, registry size $N$,
and the BoR ceiling at the current depth.

\paragraph{Action.} Binary: STOP or CONTINUE.

\paragraph{Reward.} While BoR is computed over a full test set, the RL agent needs a
per-query reward. At STOP, if the presented set contains at least one relevant tool,
the reward is $-\log_2(P_{\mathrm{rand}}(k_{\mathrm{stop}}; R_q))$, and zero otherwise.
Finding the right tool in a short list earns more reward than finding it in a long
one, because short-list success is harder to achieve by chance. A per-step
continuation cost ($\texttt{step\_cost}{=}0.01$, $\gamma{=}0.95$, except
$\gamma{=}1.0$ for MetaTool+BM25 and $\texttt{step\_cost}{=}0.005$ for BFCL+BM25)
discourages unnecessary exploration but does not encode depth preference. In the
tool-selection benchmarks studied here, the RL agent is trained with oracle $R_q$ and
assumes $R_q{=}1$ at inference, since each benchmark query has exactly one correct
tool.

\subsection{The Self-Pruning Property}
\label{sec:selfpruning}
The BoR reward naturally decreases as the shortlist grows, because $P_{\mathrm{rand}}$
rises with every additional tool shown. At $K{=}3$ from a 500-tool registry, the
reward is approximately 7 bits, but at $K{=}100$ it drops to 2 bits. The five-bit drop
occurs because random selection is far more likely to succeed at larger $K$. This
decrease is not an engineered penalty but a mathematical consequence of the metric's
structure.

\paragraph{Baselines.} We compare against two baselines. The first is Fixed-K: present
the top $K$ tools for every query regardless of difficulty. The second is an $F_1$
ablation, which tests whether any depth-aware reward produces the same adaptive
behavior. If the relevant tool is found (recall $=1$) and we present $K$ tools
(precision $=1/K$), then $F_1 = 2/(K+1)$.\footnote{The BFCL+BM25 condition (Table~\ref{tab:summary},
row 1) uses a simpler variant: a constant terminal reward of 1.0, with depth pressure
from $\gamma$-discounting alone. This is more lenient than $2/(K+1)$, making the
comparison conservative. The true $F_1$ baseline would stop earlier.} Unlike BoR, this
penalizes depth the same way regardless of the query or the registry. If depth
pressure alone were sufficient, the $F_1$ baseline should also produce adaptive
policies.

\section{Empirical Evaluation}
We evaluate BoR on three tool-selection benchmarks (BFCL, MetaTool, ToolBench) against
two baselines: Fixed-K at various depths and the $F_1$ depth penalty baseline
(Section~\ref{sec:selfpruning}). Each benchmark uses a scorer (BM25 or sentence
embeddings) to rank tools by relevance. Agent architectures are deliberately small
(DQNs and tabular Q-learning) to determine if even basic RL on CPU benefits from the
metric's structure. BoR is also validated on three document retrieval benchmarks in
Appendix~\ref{app:retrieval}. In all tool-selection experiments, $P_{\mathrm{rand}}$
is computed against the candidate set size $N$. We report ``Found\%'' as Success@K:
whether at least one relevant tool appears in the presented set. We group test queries
into difficulty buckets based on where the scorer ranks the gold tool: easy (ranked
1st), medium (ranked 2nd--5th), hard (ranked 6th--20th), and very hard (ranked 21+).

\subsection{Tool Selection}
\paragraph{BFCL (370 tools, $R_q{=}1$).} We use the Berkeley Function Calling
Leaderboard ``simple'' category, which contains 400 entries, each with one function as
ground truth. We pool all 370 unique functions into a shared registry, creating a
depth-evaluation setup not present in the original benchmark. Every other function
serves as a natural hard distractor (e.g., \texttt{calc\_area\_triangle} vs
\texttt{calculate\_triangle\_area} vs \texttt{calculate\_area}). We rank tools with
BM25 over function descriptions and parameter names. Results are reported over a
280/120 train/test split with 3 seeds. The BoR agent achieves $90.3 \pm 2.4\%$ at
$K{=}7.4 \pm 2.5$, nearly matching FK=50 ($90.8\%$) at $7\times$ less depth and
exceeding FK=20 ($87.5\%$). The $F_1$ ablation achieves $88.9 \pm 1.4\%$ at $K{=}6.4
\pm 1.9$. Both adaptive agents exceed FK=20, the strongest fixed baseline below FK=50.
The BoR agent achieves comparable coverage to FK=50 while presenting far fewer
distractors.

\paragraph{BFCL with embedding scorer.} We replace BM25 with MiniLM-L6-v2 embeddings
on the same BFCL data, 3 seeds. The stronger scorer (found@1$=73.2\%$ vs BM25's
$60.0\%$) shifts the difficulty distribution: 93 easy, 24 medium, 3 hard test queries.
The BoR agent achieves $85.0 \pm 3.0\%$ at $K{=}1.4 \pm 0.1$. The $F_1$ ablation
achieves $85.8 \pm 1.4\%$ at $K{=}1.6 \pm 0.1$. FK=5 achieves $97.5\%$. The BoR agent
adapts to the stronger scorer by stopping much earlier ($K{=}1.4$ vs $K{=}7.4$ with
BM25). The same reward produces different stopping behavior because the stronger
scorer ranks the correct tool earlier and changes the score distribution the agent
sees.

\paragraph{MetaTool with embedding scorer (199 tools, $R_q{=}1$).} We replace BM25
with all-MiniLM-L6-v2 embeddings on 2{,}000 queries (1{,}400 train / 600 test), using
100-tool candidate sets constructed from the gold tool, 24 hard distractors, and 75
randomly sampled tools from the registry. The BoR agent achieves $K{=}2.3$,
found$=73.3\%$, BoR$=4.44$ bits. The $F_1$ ablation achieves $K{=}3.0$, found$=69.0\%$,
BoR$=4.24$ bits. The BoR agent exceeds the ablation on both coverage and selectivity
while stopping earlier.

\paragraph{MetaTool with varying candidate-set size ($N{=}20, 50, 100$).} Using the
same queries and embedding scorer, we vary the candidate-set size while constructing
each set from the gold tool plus the top hard distractors, with no random fill. We use
1{,}200 queries (840 train / 360 test). The BoR agent's reward increases with $N$
(2.75 bits at $N{=}20$, 3.66 at $N{=}50$, 4.36 at $N{=}100$), because finding the
correct tool in a larger candidate set is worth more than finding it in a smaller one.
By contrast, the $F_1$ ablation stays near 63--64\% found and Fixed $K{=}1$ stays near
59\%, showing little sensitivity to $N$.

The difficulty-bucket analysis at $N{=}50$ shows how the adaptation works. Because the
candidate sets are smaller, we use narrower buckets: ranked 1st, 2nd--3rd, 4th--10th,
and 11+. On easy queries (gold ranked 1st), all methods achieve 100\% found, with the
BoR agent stopping at $K{=}1.58$. On medium queries (gold ranked 2nd--3rd), the BoR
agent extends to $K{=}3.72$ and finds $79.3\%$, while the $F_1$ ablation at $K{=}1.48$
finds only $29.3\%$ and Fixed $K{=}1$ finds 0\%. On hard queries (gold ranked
4th--10th), the BoR agent increases to $K{=}4.24$ and finds $26.5\%$, while the
ablation and Fixed $K{=}1$ find nothing.

\paragraph{ToolBench (3{,}251 tools, tool-level matching).} We use the official
ToolBench data~\citep{qin2024toolllm}, matching at the tool level rather than the API
level.\footnote{We use the full G1 single-tool queries from the official Google Drive
distribution~\citep{qin2024toolllm}, yielding 3{,}251 tools.} Because each tool exposes
multiple APIs, we aggregate API descriptions to the tool level and score them with
MiniLM embeddings. We use 2{,}000 single-tool queries (1{,}400 train / 600 test) and
construct $N{=}50$ candidate sets from the gold tool plus hard distractors, with 3
seeds. The BoR agent achieves $61.9 \pm 0.6\%$ at $K{=}4.4 \pm 0.4$. The $F_1$ ablation
achieves $47.6 \pm 1.3\%$ at $K{=}1.5 \pm 0.3$, barely above FK=1 ($45.3\%$). A common
fixed-depth baseline, FK=5, achieves $64.7\%$, while deeper baselines reach $77.3\%$
(FK=20) at the cost of selectivity.

The difficulty-bucket analysis at $N{=}50$ shows the clearest separation between
methods, with 3-seed standard deviations indicating that the pattern is stable. On
easy queries (gold ranked 1st, $n{=}272$), the BoR agent uses $K{=}2.5 \pm 0.2$ and
achieves 100\%. On medium queries (gold ranked 2nd--5th, $n{=}116$), it increases to
$K{=}4.8 \pm 0.5$ and finds $74.4 \pm 0.4\%$. FK=5 also finds 100\% on medium queries,
but only by using the same depth for every query. On hard queries (gold ranked
6th--20th, $n{=}76$), the BoR agent increases to $K{=}5.7 \pm 0.5$ and finds $16.7 \pm
4.3\%$, while FK=5, the $F_1$ ablation, and FK=1 all find 0\%. On very hard queries
(gold ranked 21+, $n{=}136$), the BoR agent increases to $K{=}6.9 \pm 0.7$ and finds
$0.2\%$, while all alternatives find 0\%. Overall, the $F_1$ ablation finds only
$47.6\%$.

\begin{figure}[t]
\centering
\includegraphics[width=\textwidth]{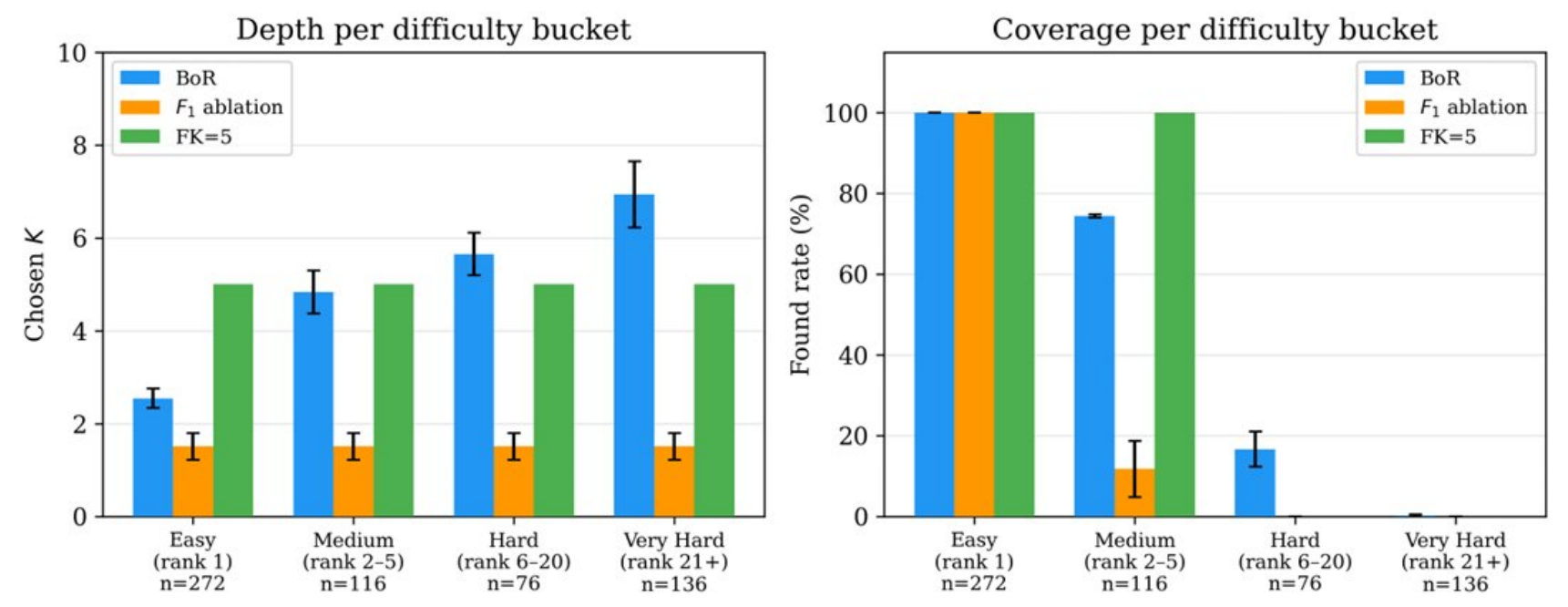}
\caption{ToolBench difficulty buckets at $N{=}50$. Left: chosen $K$ per difficulty
bucket for BoR, the $F_1$ baseline, and FK=5. Right: found rate per bucket. The BoR
agent adapts $K$ from 2.5 (easy) to 6.9 (very hard), while the $F_1$ baseline stays
near 1.5 and FK=5 stays fixed at 5.}
\label{fig:toolbench}
\end{figure}

The BoR agent adapts $K$ from 2.5 to 6.9 based on query difficulty
(Figure~\ref{fig:toolbench}). By contrast, the $F_1$ ablation stays near $K{\approx}1.5$
across all buckets, while Fixed $K{=}5$ uses $K{=}5$ everywhere. On hard queries, the
BoR agent finds tools that every alternative misses entirely. FK=5 achieves higher
aggregate coverage ($64.7\%$ vs $61.9\%$). This reflects its uniform depth: it catches
all easy and medium queries (where gold is ranked 1--5) but finds nothing on hard and
very hard queries (where gold is ranked 6+). The BoR agent trades some medium-query
coverage for recovery on hard queries.

\subsection{Effect of Scorer Quality}
\paragraph{MetaTool with BM25 scorer.} On MetaTool, BM25 achieves only 33\% found@1.
Tool descriptions use names like \texttt{reflect\_notes} for queries about ``time
dilation'', so there is little lexical overlap for BM25 to exploit. With such weak
score signal, the BoR agent expands to $K{=}80.7$, nearly showing all tools. It
achieves a high found rate ($96.2\%$) but only 1.04 bits of selectivity. The $F_1$
baseline stops earlier ($K{=}57.2$) and achieves $82.8\%$ found.

This is a negative result, but an informative one. At $K{=}50$, the BoR ceiling is
still 1.0 bit, so the RL agent continues. Without a discriminative scorer, there is no
reliable signal for where to stop. The $F_1$ baseline, by contrast, stops early
regardless of scorer quality, so the weak scorer is effectively hidden.

\begin{figure}[t]
\centering
\includegraphics[width=\textwidth]{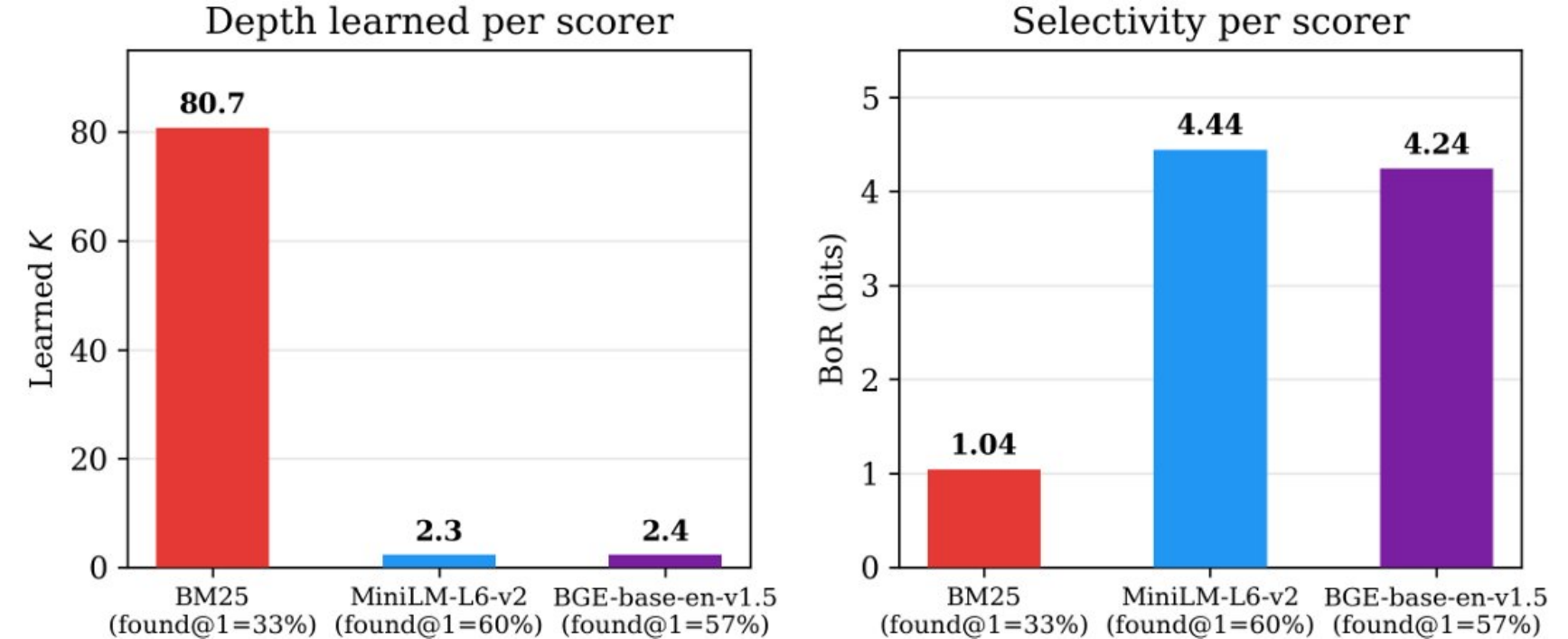}
\caption{Scorer ablation on MetaTool. Three scorers (BM25, MiniLM, BGE) on the same
data with the same BoR reward produce different learned depths. BM25 leads to
$K{=}80.7$, while both embedding scorers lead to $K{\approx}2.3$.}
\label{fig:scorer}
\end{figure}

\paragraph{Scorer comparison.} On the same MetaTool data (Figure~\ref{fig:scorer})
with the same BoR reward, three scorers produce different learned depths: BM25
(found@1$=33\%$) leads to $K{=}80.7$, MiniLM-L6-v2 (found@1$=60\%$) leads to
$K{=}2.3$, and BGE-base-en-v1.5 (found@1$=57\%$) leads to $K{=}2.4$. The two embedding
models, which have similar retrieval quality, produce nearly identical policies. The
difference is between lexical and semantic scoring, not between embedding
architectures.

\subsection{Downstream Tool-Choice Validation}
\label{sec:downstream}
The experiments in Sections 4.1 and 4.2 evaluate a necessary condition: whether the
correct tool appears in the presented set. A natural follow-up is whether adaptive
depth also helps the LLM select the correct tool from that set. We test this on BFCL
using the full 370-function registry and the same BM25 scorer as in Section 4.1. The
RL agents are retrained for this setup with the standard hyperparameters
($\texttt{step\_cost}{=}0.01$, $\gamma{=}0.95$), which produces shorter lists
($K{=}2.2$) than the main-table agents ($K{=}7.4$, trained with
$\texttt{step\_cost}{=}0.005$). Each method's tool set is presented to Claude Sonnet
4.6 with a constraint that forces the model to select exactly one tool. Tool
descriptions and ordering from the scorer are kept identical across methods, with only
the set size varying. We evaluate on 120 test queries with $N{=}370$, 3 seeds, and
temperature$=0$.

\begin{table}[t]
\centering
\caption{Downstream tool-choice validation on BFCL with Claude Sonnet 4.6.
Presented\% is the fraction of queries where the gold tool appears in the presented
set. Choice Acc\% is the fraction of presented-gold queries where Claude selects the
gold tool. End-to-End\% is the product. Values after $\pm$ are standard deviations
across 3 seeds. FK=5 and FK=1 are deterministic (no seed variance).}
\label{tab:downstream}
\begin{tabular}{lcccc}
\toprule
Method & Presented\% & Choice Acc\% & End-to-End\% & Avg $K$ \\
\midrule
BoR  & $76.9 \pm 0.4$ & $93.1 \pm 0.5$ & $71.7 \pm 0.0$ & $2.2 \pm 0.4$ \\
$F_1$ & $72.8 \pm 2.7$ & $94.3 \pm 1.7$ & $68.6 \pm 1.7$ & $1.7 \pm 0.3$ \\
FK=5 & $84.2$ & $87.1$ & $73.3$ & $5.0$ \\
FK=1 & $65.0$ & $100.0$ & $65.0$ & $1.0$ \\
\bottomrule
\end{tabular}
\end{table}

The main finding (Table~\ref{tab:downstream}) is that over-presentation reduces
downstream choice accuracy. When the gold tool is present, Claude selects it $93.1 \pm
0.5\%$ of the time under BoR ($K{=}2.2$) but only $87.1\%$ under FK=5, a gap of 6
percentage points. FK=1 achieves 100\% choice accuracy because a single tool leaves no
room for confusion, but its low presentation rate ($65.0\%$) limits end-to-end
accuracy.

The per-bucket breakdown shows the effect most clearly on medium-difficulty queries
(gold ranked 2nd through 5th, $n{=}23$). FK=5 presents the gold tool on every query
but Claude selects it only $60.9\%$ of the time. The BoR agent presents on $62.3 \pm
2.0\%$ of queries, but when it does, Claude selects correctly $76.8 \pm 2.5\%$ of the
time. FK=1 never presents on medium queries (0\% end-to-end). On easy queries (gold
ranked 1st, $n{=}78$), all methods perform well once the gold tool is presented, but
the highest end-to-end accuracy comes from the shortest lists: FK=1 reaches $100.0\%$,
$F_1$ $97.0\%$, BoR $96.2\%$, and FK=5 $94.9\%$.

FK=5 achieves the highest end-to-end accuracy ($73.3\%$) because most BFCL queries are
easy, so simply showing more tools ensures the gold tool is present. The BoR agent
reaches $71.7 \pm 0.0\%$, close behind despite presenting the gold tool on fewer
medium queries. The $F_1$ ablation trails at $68.6 \pm 1.7\%$. On very hard queries
(gold ranked 21+), the BoR agent searches deeper ($K{=}5.9$), attempting recovery when
the scorer provides weak signal. This is consistent with the ToolBench difficulty-
bucket results above.

We replicate this experiment with an embedding scorer (MiniLM-L6-v2) and observe the
same pattern with a wider gap: $96.1 \pm 2.0\%$ choice accuracy for BoR versus $84.6\%$
for FK=5 (11.5pp), rising to $80.1\%$ versus $50.0\%$ on medium queries (30pp). The
effect is consistent across scorer types. The embedding scorer's higher found@1 on
this split (77\% vs 65\%) shifts the difficulty distribution toward easy queries,
which compresses the aggregate gap, but the per-bucket pattern is identical: adaptive
depth improves choice accuracy when the gold tool is present.

These results indicate that adaptive depth not only presents the right tool more
efficiently but also improves the LLM's ability to choose it by reducing distractor
load.

\section{Discussion}

\subsection{Cross-Condition Patterns}
Table~\ref{tab:summary} summarizes all nine tool-selection conditions. Three
observations stand out. First, the BoR agent often achieves competitive found rates at
substantially lower depth. In some conditions it nearly matches the best Fixed-K
baseline (e.g., BFCL+BM25), while in others a fixed depth achieves higher aggregate
coverage when its uniform $K$ happens to include the gold tool more often (e.g., FK=5
at $97.5\%$ vs BoR at $85.0\%$ on BFCL+embed, FK=5 at $64.7\%$ vs BoR at $61.9\%$ on
ToolBench). Second, the same BoR reward produces different policies depending on the
scorer and the registry: $K{=}7.4$ with BM25 versus $K{=}1.4$ with embeddings on BFCL,
and $K{=}80.7$ versus $K{=}2.3$ on MetaTool. No tuning was changed between conditions.
Third, unlike BoR, the $F_1$ baseline remains shallow in the candidate-set-size
experiments and does not increase depth in response to harder query buckets,
confirming that its pressure comes from a fixed depth penalty rather than chance
correction.

\begin{table}[t]
\centering
\caption{Summary across nine tool-selection conditions. Full registry sizes:
BFCL$=370$, MetaTool$=199$, ToolBench$=3{,}251$. In condition names, $N$ denotes
candidate set size used for evaluation.}
\label{tab:summary}
\small
\begin{tabular}{lccccc}
\toprule
Condition & BoR Found\% & $F_1$ Found\% & Best FK Found\% & BoR $K$ & $F_1$ $K$ \\
\midrule
BFCL+BM25 (N=370)      & $90.3 \pm 2.4$ & $88.9 \pm 1.4$ & 90.8 (FK=50) & $7.4 \pm 2.5$ & $6.4 \pm 1.9$ \\
BFCL+embed (N=370)     & $85.0 \pm 3.0$ & $85.8 \pm 1.4$ & 97.5 (FK=5)  & $1.4 \pm 0.1$ & $1.6 \pm 0.1$ \\
MetaTool+BM25 (N=100)  & 96.2          & 82.8          & 83.7 (FK=50) & 80.7         & 57.2 \\
MetaTool+embed (N=100) & 73.3          & 69.0          & 74.7 (FK=3)  & 2.3          & 3.0  \\
MetaTool+BGE (N=100)   & 71.4          & 61.4          & 57.2 (FK=1)  & 2.4          & 1.2  \\
MetaTool (N=20)        & 70.6          & 63.3          & 59.2 (FK=1)  & 1.6          & 1.3  \\
MetaTool (N=50)        & 74.7          & 63.9          & 59.2 (FK=1)  & 2.6          & 1.2  \\
MetaTool (N=100)       & 72.8          & 63.3          & 59.2 (FK=1)  & 2.1          & 1.5  \\
ToolBench (N=50)       & $61.9 \pm 0.6$ & $47.6 \pm 1.3$ & 77.3 (FK=20) & $4.4 \pm 0.4$ & $1.5 \pm 0.3$ \\
\bottomrule
\end{tabular}
\end{table}

\subsection{Scope and Transferability}
The framework transfers to document retrieval (Appendix~\ref{app:retrieval}) because
the combinatorial structure is the same: select $K$ from $N$ where $R_q$ are relevant.
In tool selection, extra candidates can actively hurt LLM accuracy. In document
retrieval, the cost is mainly wasted depth. BoR measures selectivity in both cases.

None of the benchmarks we use were designed to evaluate search depth. Our evaluation
code, included in the supplementary material, downloads each benchmark's public data
and constructs candidate sets at varying $K$ for depth evaluation. This works but has
limitations: BFCL's curated function definitions make most queries easy for embedding
scorers, while ToolBench's noisy aggregated descriptions produce a wider difficulty
range. Neither provides native ground truth for the question ``how many tools should
the LLM see?'' A purpose-built benchmark with controlled difficulty distributions
would benefit future work.

\subsection{Limitations}
The empirical setup is intentionally lightweight. Our RL agents are small DQNs or
tabular Q-learning models rather than production-scale policies, because the goal is to
test whether the reward itself induces adaptive depth. MetaTool results are reported
from single-seed runs, although they cover six conditions. BFCL and ToolBench include
3-seed standard deviations. These choices limit the absolute scale of the experiments
but they do not change the main qualitative finding: the same BoR reward produces
adaptive depth behavior across benchmarks, scorers, and registry sizes. More
informative tool descriptions appear to support a wider range of adaptive depths,
though recovery on very hard queries remains limited ($0.2\%$ on ToolBench).

Our evaluation targets whether the correct tool appears in the presented set.
Downstream tool choice accuracy is validated on BFCL, but whether the LLM calls the
tool with the right arguments is outside the scope of this work.

\section{Conclusion}
We treated the number of tools shown to an LLM agent as the object of evaluation. We
applied the BoR metric to evaluate and control search depth across multiple
tool-selection conditions and document retrieval benchmarks. The metric adapts to
registry size, scorer quality, and query difficulty without per-condition tuning. When
used as an RL reward, it produces query-dependent depth policies, unlike Fixed-K or
the $F_1$ baseline. Downstream validation on BFCL indicates that the resulting shorter
lists improve the LLM's tool-choice accuracy. The main limitation is that BoR
optimizes selectivity rather than maximum recall, so Fixed-K can achieve higher
aggregate coverage when a uniform depth happens to suffice.

\appendix
\section{Retrieval Validation}
\label{app:retrieval}
We validate BoR on three document retrieval benchmarks with the same mathematical
structure (select $K$ from $N$ where $R_q$ are relevant) and well-understood datasets
(Table~\ref{tab:retrieval}). This suggests that the metric and the self-pruning
property transfer beyond tool selection.

\paragraph{SciFact ($N{=}5{,}183$, $R_q \approx 1$).} We use a real BM25 retriever
with tabular Q-learning on 210 train / 90 test queries. The BoR agent learns $K{=}7.2$
on average (std$=3.34$) and achieves $78.9\%$ found rate, placing between FK=10
($75.6\%$) and FK=20 ($81.1\%$) at $7\times$ less depth than FK=50 ($85.6\%$). The
$F_1$ ablation learns $K{=}5.0$ with $66.7\%$ found rate and $K$ std$=0.00$, showing no
adaptation at all.

\paragraph{NFCorpus ($N{=}3{,}633$, $R_q{=}1$--$475$).} Real BM25 retriever with a
DQN. 226 train / 97 test queries. The BoR agent achieves $K{=}22.9$, found$=71.1\%$
with $K$ std$=30.53$. The $F_1$ ablation achieves $K{=}24.5$, found$=68.0\%$ with $K$
std$=31.92$. Unlike the tool-selection settings, both RL agents show adaptive depth on
this dataset, and the separation is directional rather than dramatic. The per-$R_q$
breakdown shows where policies diverge: on dense queries ($R_q>30$), the BoR agent uses
$K{=}10.3$ versus $F_1$'s $K{=}13.7$, achieving the same found rate at 25\% less depth.

\paragraph{MS MARCO ($N{=}8.8$M).} Real BM25 retriever using \texttt{ir\_datasets}
with a DQN. 350 train / 150 test queries. The local BM25 index contains about 51K
passages, but BoR is computed against the full corpus ($N{=}8{,}841{,}823$). The BoR
agent achieves $K{=}24.0$, found$=82.7\%$, exceeding FK=50's found rate ($80.7\%$) at
less than half the depth.

\begin{table}[h]
\centering
\caption{Retrieval validation across three document retrieval benchmarks.}
\label{tab:retrieval}
\small
\begin{tabular}{lccccccc}
\toprule
Condition & $N$ & BoR Found\% & $F_1$ Found\% & Best FK Found\% & BoR $K$ & $F_1$ $K$ & Reward (bits) \\
\midrule
SciFact  & 5{,}183 & 78.9 & 66.7 & 85.6 (FK=50) & 7.2  & 5.0  & 9.76 \\
NFCorpus & 3{,}633 & 71.1 & 68.0 & 69.1 (FK=50) & 22.9 & 24.5 & 4.90 \\
MS MARCO & 8.8M   & 82.7 & 78.7 & 80.7 (FK=50) & 24.0 & 16.8 & 20.28 \\
\bottomrule
\end{tabular}
\end{table}

Across all three corpora, one reward function produces three different learned
policies. The BoR agent adapts its depth to corpus structure without per-corpus tuning.

\end{document}